\documentclass[aps,prd,reprint,superscriptaddress,preprintnumbers,longbibliography]{revtex4-1}

\usepackage{amsmath,amssymb,bm,mathrsfs,bbold,dsfont}
\usepackage{xfrac}
\usepackage[dvipsnames]{xcolor}
\usepackage{physics}
\usepackage{slashed}
\usepackage{multirow}
\usepackage{booktabs}

\usepackage{epsfig}
\usepackage{psfrag}
\usepackage{tikz}
\usepackage{tikz-feynman}
\usepackage{float}
\usepackage[caption=false]{subfig}  % Prevents conflict with revtex4-1
\PassOptionsToPackage{caption=false}{subfig}

\usepackage[sort&compress]{natbib}
\usepackage[colorlinks=true,
    linkcolor=red,
    citecolor=blue,
    urlcolor=blue,
    filecolor=blue,
    anchorcolor=blue,
    menucolor=blue,
    linktocpage=true,
    pdfproducer=medialab,
    pdfa=true
]{hyperref}
\usepackage{cleveref}

\usepackage{setspace}
\usepackage{relsize}
\usepackage{enumerate}
\usepackage{comment}
\usepackage[normalem]{ulem}
\usepackage{srcltx}
\usepackage{xcolor}
\usepackage{diagbox}
\usepackage{appendix}

\newcommand{\emphb}[1]{\textbf{\emph{#1}}}

\newcommand{\TRH}{T_{\rm RH}}

\newcommand{\mKK}{ m_{\rm KK}}
\newcommand{\GeV}{{\rm GeV}}
\newcommand{\MeV}{{\rm MeV}}

\newcommand{\eV}{{\rm eV}}
\newcommand{\meV}{{\rm meV}}

\begin{document}

\newcommand{\ourtitle}{\boldmath Near Threshold Kaluza-Klein Graviton Decays and the Dark Dimension}

\title{\ourtitle}

\author{Kevin Langhoff}
\email{langhoff@mit.edu}
\affiliation{Center for Theoretical Physics -- a Leinweber Institute, Massachusetts Institute of Technology, Cambridge, MA 02139, USA}

\begin{abstract}\noindent
Near threshold, intra-tower decays of Kaluza-Klein (KK) gravitons are pure $d$-wave, with rate $\Gamma \propto p^5$ in the decay products momentum $p$, rather than the $s$-wave $\Gamma \propto p$ assumed in the dark dimension literature. In a flat dark dimension this excludes KK gravitons as a dark matter candidate. Even if some other source of dark matter is assumed, the irreducible freeze-in abundance of KK gravitons produce MeV gamma rays which allows one to set upper limits on the reheating temperature; additionally using the lower bound on the reheating temperature from big-bang nucleosynthesis and the cosmic microwave background requires the radius of the extra dimension to be almost three orders of magnitude smaller than the distance conjecture would naively prefer, $R\sim \Lambda^{-1/4}$. 
\end{abstract}
\maketitle

\section{Introduction}\label{sec:Introduction}

Recently it has been pointed out that swampland principles \cite{Ooguri:2006in,Lee:2019wij,Lee:2019xtm,Lust:2019zwm,Obied:2018sgi,Bedroya:2019snp} together with the observed cosmological constant, $\Lambda^{1/4}\approx 2.3~{\rm meV}$, hint at the possible existence of a single mesoscopic extra dimension with a size of several microns  \cite{Montero:2022prj,Vafa:2024fpx} called the \emph{Dark Dimension}. Within this framework one obtains a tower of Kaluza-Klein (KK) graviton excitations with masses $m_n\approx n\,\mKK$, $\mKK=1/R\sim\Lambda^{1/4}$ for a radius $R$, which can source the dark matter (DM) \cite{Gonzalo:2022jac}. The tower is unavoidably produced via UV freeze-in by Planck-suppressed interactions~\cite{Giudice:1998ck, Hall:1999mk}
\begin{align}
    \label{eq:KK_graviton_coupling}
    \mathcal{L} \supset -\frac{c_n}{M_P}h^{(n)}_{\mu \nu} T^{\mu \nu}\,,
\end{align}
where $M_P$ is the 4D reduced Planck mass, $T^{\mu\nu}$ the SM stress tensor, $n$ labels the KK mode and $c_n=\sqrt{L}\,\psi_n(y_b)$ is its profile at the Standard Model (SM) brane, $y=y_b$, with $\psi_n$ and $L$ defined in Sec.~\ref{sec:decay}.

In the initial construction, with a flat extra dimension and a uniformly spaced tower, freeze-in generates the observed relic abundance if the reheating temperature is $\TRH \sim \mathcal{O}({1\,\rm GeV})$~\cite{Gonzalo:2022jac}, thanks to the very large number of accessible modes, $\TRH/\mKK\sim \mathcal{O}(10^{11})$.

At production, the energy density is dominated by modes with masses near $\TRH$, which decay to SM particles via Eq.~\eqref{eq:KK_graviton_coupling} with lifetimes $\tau \sim \mathcal{O}(10^{5}~{\rm yrs})$. To survive as DM, these heavy modes must instead \emph{cascade} down to lower KK levels through intra-tower decays~\cite{Mohapatra:2003ah,Dienes:2011ja,Dienes:2011sa,Dienes:2012jb}, leaving a present-day mass distribution near the keV scale; such decays \emph{are expected} to outpace decays to the SM, given the number of open channels~\cite{Gonzalo:2022jac}.

Let us briefly \emph{assume} the cascade is fast enough to move the tower's mass distribution to lighter masses. Two bounds then pull in opposite directions. The decay width of the gravitons to SM photons grows as $m^3$ and energy injection into the CMB~\cite{Slatyer:2016qyl} requires the surviving modes to be light, near the keV scale~\cite{Gonzalo:2022jac}. However, if the masses become too light, DM decays give the decay product particles too much kinetic energy and DM becomes too warm to remain in the observed DM halos~\cite{Obied:2023clp}. Decays must therefore be \emph{near threshold} and release as little kinetic energy as possible, i.e. be nearly mass conserving (or equivalently KK number conserving for a flat extra dimension).

In this letter we point out that the above tension is far more severe than previously claimed. Near threshold, KK graviton decays are in fact $d$-wave suppressed~\cite{deGiorgi:2021xvm}, not $s$-wave as assumed in the cascade estimates of Refs.~\cite{Gonzalo:2022jac,Obied:2023clp}: the rate scales with the final state momentum as $p^5$ rather than $p$. For their flat dark dimension, the initial population, with masses of a few $\TRH$, then decays at roughly $(\mKK/m)^2\approx10^{-24}$ of the $s$-wave rate at the same coupling for $m=1\,\GeV$ and $\mKK=1\,\meV$. 

Even if KK gravitons are not dark matter, an irreducible population is still produced via freeze-in (analogous to~\cite{Langhoff:2022bij}) and their decays into MeV photons bound $\TRH$ from above. Additionally, $\TRH$ is bounded from below by big-bang nucleosynthesis (BBN) and the cosmic microwave background (CMB). This excludes the existence of a flat dark dimension with $R\gtrsim0.2\,\mu{\rm m}$ (Fig.~\ref{fig:window}). 

Note Added: Ref.~\cite{Lee:2026myr}, which appeared while this Letter was being completed, obtains the same threshold suppression. Our result was obtained independently and by a different route. Sec.~\ref{sec:decay} adds the double-copy form of the vertex, a proof that the decay is pure $d$-wave, and a different reading of the cancellation. We also calculate a bound on $\TRH$ from the irreducible population of gravitons produced via freeze-in (for large extra dimensions see Refs.~\cite{Hall:1999mk,Hannestad:2001nq}).

\section{The Decay Rate}\label{sec:decay}

In this section we review the calculation for the decay of KK gravitons and emphasize the relevant limit for dark dimension scenarios where decays are assumed to be nearly mass conserving. We show that $d$-wave suppression of the amplitude causes this decay rate to be greatly suppressed relative to the $s$-wave rate often assumed in the dark dimension literature.

\emphb{The Action and Interactions.}  We consider five-dimensional minimally coupled gravity with action
\begin{align}\label{eq:S}
    S = \frac{M_5^3}{2}\int d^5x\sqrt{-G}R_5 + S_m,
\end{align}
where $M_5$ is the reduced 5D Planck mass and $S_m$ contains bulk scalars with arbitrary potentials, brane tensions, and the Gibbons--Hawking terms at the branes. The coordinate $y$ runs over the interval $0\le y\le\pi R$ and primes denote derivatives with respect to $y$. The background metric is
\begin{align}\label{eq:ds2}
    ds^2 = e^{2A(y)}\eta_{\mu \nu}dx^\mu dx^\nu + dy^2,
\end{align}
with signature $(-,+,+,+,+)$. Gravitons are tensor fluctuations parametrized as $G_{\mu \nu } = e^{2A}(\eta_{\mu\nu} + h_{\mu\nu})$. We expand the graviton into modes as
\begin{align}\label{eq:modes}
    h^{\mu \nu}(x,y) = \frac{2}{M_{5}^{3/2}}\sum_n \psi_n(y) h_n^{\mu\nu}(x),
\end{align}
with $\psi_n(y)$ which solve the Sturm--Liouville problem
\begin{align}\label{eq:SL}
    -(e^{4A}\psi_n')' = m_n^2 e^{2A} \psi_n
\end{align}
with Neumann boundary conditions and normalization
\begin{align}
    \int dy\, e^{2A}\psi_m(y) \psi_n(y) = \delta_{mn}.
\end{align}
The zero mode is $\psi_0 = 1/\sqrt{L}$ with $L \equiv \int dy\, e^{2A}$, and $M_P^2=M_5^3L$.

At quadratic order, for the physical transverse-traceless polarizations ($\partial_\mu h_n^{\mu\nu}=\eta_{\mu\nu}h_n^{\mu\nu}=0$), we have
\begin{align}
    \mathcal{L}_2 = -\frac{1}{2}(\partial_\alpha h_n^{\mu \nu})^2 - \frac{1}{2}m_n^2 (h_n^{\mu \nu})^2.
\end{align}
where the mass of the $n^{\rm th}$ mode is $m_n^2=\int dy\,e^{4A}\psi_n'^2$.

At third order and for three distinct modes we have 
\begin{align}\label{eq:L3}
    \mathcal{L}_3 = \frac{4g_{mn\ell}}{M_P}\big[\mathcal{L}_V + \mathcal{L}_\Box + \mathcal{L}_m\big],
\end{align}
where $\mathcal{L}_V+\mathcal{L}_\Box$ is the cubic part of the 4D Einstein--Hilbert term $\sqrt{-g}R$ with the replacement $g=\eta+h_m+h_n+h_\ell$ and is split into terms where every derivative is contracted with a field, $\mathcal{L}_V$, and terms where two derivatives are contracted with each other,
\begin{align}
    \mathcal{L}_\Box &\equiv -\frac{1}{2}\Big[\Tr(\Box h_m h_n h_\ell)+\Tr(h_m\Box h_n h_\ell)\notag\\
    &\qquad\quad+\Tr(h_m h_n \Box h_\ell)\Big],\label{eq:LBox}
\end{align}
Additionally, there is a term proportional to the sum of the graviton masses, $m_\Sigma^2 = m_m^2 + m_n^2 + m_\ell^2$, which is
\begin{align}
    \mathcal{L}_m &\equiv \frac{1}{2} m_\Sigma^2\Tr(h_m h_n h_\ell). \label{eq:Lm}
\end{align}
The dimensionless couplings $g_{mn\ell}$ come from the overlaps
\begin{align}\label{eq:g}
    g_{mn\ell} = \psi_0^{-1}\int dy\, e^{2A}\psi_m \psi_n \psi_\ell.
\end{align}
In a flat dimension, $\psi_n=\sqrt{2/(\pi R)}\cos(ny/R)$ for $n\ge1$ and $g_{mn\ell}=\delta_{m,n+\ell}/\sqrt2$ for $n,\ell<m$: every allowed channel sits exactly at threshold, and any decay requires KK-number violation from a $y$-dependent background~\cite{Mohapatra:2003ah}.

The term in Eq.~\eqref{eq:Lm} comes from the gradient energy along $y$ in Eq.~\eqref{eq:S}, whose cubic part $(1/2) e^{4A}\Tr(h h' h')$ has overlaps $b_{rs;t}=\int dy\, e^{4A}\psi_r'\psi_s'\psi_t$. Integrating by parts with Eq.~\eqref{eq:SL} gives,
\begin{align}\label{eq:SLid}
    b_{rs;t}=\frac{1}{2}\big(m_r^2+m_s^2-m_t^2\big)\,\int dy\,e^{2A}\psi_r\psi_s\psi_t.
\end{align}
Summing over undifferentiated legs gives Eq.~\eqref{eq:Lm}.

For massless gravitons $\mathcal{L}_\Box$ vanishes on shell and leaves only $\mathcal{L}_V$. For massive on shell KK gravitons, we have $\Box h_i = m_i^2 h_i$, such that $\mathcal{L}_\Box=-\mathcal{L}_m$ and
\begin{align}\label{eq:L3V}
    \mathcal{L}_3 \to \frac{4g_{mn\ell}}{M_P}\,\mathcal{L}_V.
\end{align}

\emphb{The 3-Point Amplitude.} The on-shell 3-point amplitude for graviton $h_m$ (particle 1) to gravitons $h_n$ and $h_\ell$ (particles 2 and 3), with all ingoing momenta $q_1 + q_2 + q_3 = 0$, is $\mathcal{M} = (4g_{mn\ell}/M_P)\,V$ with 
\begin{align}
    &V = \frac{1}{8} \varepsilon_1^{\mu \nu} \varepsilon_2^{\rho \sigma} \varepsilon_3^{\alpha \beta} Y_{\mu \rho \alpha} Y_{\nu \sigma \beta}\\
    &Y_{\mu \rho \alpha} \equiv \eta_{\mu \rho }(q_1 - q_2)_\alpha + \eta_{ \rho \alpha } (q_2 - q_3)_\mu\notag\\
    &\qquad\quad+  \eta_{ \alpha \mu}  (q_3 - q_1)_\rho.
\end{align}
Here $Y$ is the color-stripped cubic Yang--Mills vertex; the on-shell three-graviton amplitude is its square (i.e. the double copy~\cite{Kawai:1985xq,Bern:2008qj,Bern:2019prr}). Therefore, $V$ is the massless three-graviton amplitude evaluated with massive momenta.

Importantly, the amplitude vanishes at threshold ($m_m = m_n + m_\ell$), where every momentum has only a time component while all polarizations are purely spatial; since every $q_i$ in $Y$ is contracted with a polarization tensor and vanishes the amplitude is zero in this limit. This is not rotational symmetry, which allows the $s$-wave vertex built from $\Tr(\varepsilon_1\varepsilon_2\varepsilon_3)$. In other words, the zero is dynamical in origin.

To calculate the total decay rate off threshold we sum over polarizations with
\begin{align}\label{eq:polsum}
    \sum_{\lambda}\varepsilon_\lambda^{\mu\nu}\varepsilon_\lambda^{*\rho\sigma}
    =\tfrac12\big(\Pi^{\mu\rho}\Pi^{\nu\sigma}+\Pi^{\mu\sigma}\Pi^{\nu\rho}\big)-\tfrac13\Pi^{\mu\nu}\Pi^{\rho\sigma},
\end{align}
where $\Pi^{\mu\nu}=\eta^{\mu\nu}+q^\mu q^\nu/m^2$ for a graviton of momentum $q$ and mass $m$. Using this and defining the dimensionless mass ratio $r_{n,\ell}\equiv m_{n,\ell}/m_m$, we obtain
\begin{align}
    &\sum_{\rm pol}|V|^2 = \frac{p^4\,F(r_n,r_\ell)}{216\, r_n^4 r_\ell^4},\\
    &F(r_n,r_\ell) = s_1^4 + 22\,s_1^2 s_2 + 76\,s_2^2 + 90\,s_1 s_3,
\end{align}
with $s_1=1+r_n^2+r_\ell^2$, $s_2=r_n^2+r_\ell^2+r_n^2r_\ell^2$, $s_3=r_n^2r_\ell^2$, and $p$ the magnitude of either daughter's 3-momentum in the parent rest frame. With $\Gamma=(p/8\pi m_m^2)\tfrac15\sum_{\rm pol}|\mathcal M|^2$ we obtain, for distinct gravitons,
\begin{align}\label{eq:Gexact}
    \Gamma(h_m\to h_nh_\ell)=\frac{g_{mn\ell}^2}{540\pi}\,\frac{p^5}{m_m^2\,M_P^2}\,\frac{F(r_n,r_\ell)}{r_n^4\,r_\ell^4},
\end{align}
with an additional factor of $\frac{1}{2}$ if $n=\ell$; this is in agreement with the widths of Refs.~\cite{deGiorgi:2021xvm,Chivukula:2024nzt,Lee:2026myr}. We can simplify this further by defining the kinetic energy released in a decay as $Q \equiv m_m - (m_n+m_\ell)$ and the reduced mass of the final state, $\mu_{n\ell} = m_nm_\ell/(m_n+m_\ell)$, such that
\begin{align}\label{eq:Gthr}
    &\Gamma(h_m\to h_nh_\ell) \approx \frac{g_{mn\ell}^2}{3\pi }\frac{ m_m^2\, p^5}{\mu_{n\ell}^4 M_P^2}f(\mu_{n\ell}/m_m)\\
    &f(x) \equiv (1-x)(1-3x+4x^2-x^3)
\end{align}
where we take the limit where $F$ is taken to be its value on threshold which is valid for $Q\ll m_{n,\ell}$.  
With $p^2\approx 2\mu_{n\ell}Q$ this suppression is $\propto Q^{5/2}$. Had the mass term in Eq.~\eqref{eq:L3} survived on shell, it would give a threshold amplitude $V_s=\tfrac12 m_\Sigma^2\Tr(\varepsilon_1\varepsilon_2\varepsilon_3)$ and $s$-wave width $\Gamma_s=7g_{mn\ell}^2m_m^2p\,(1-\mu_{n\ell}/m_m)^2/(6\pi M_P^2)$. Summing over $\delta n$ KK levels gives the scaling assumed in Ref.~\cite{Gonzalo:2022jac}. The ratio to this value is
\begin{align}\label{eq:ratio}
    \frac{\Gamma}{\Gamma_s}=\frac{8}{7}\,\frac{f(\mu_{n\ell}/m_m)}{(1-\mu_{n\ell}/m_m)^2}\Big(\frac{Q}{\mu_{n\ell}}\Big)^2.
\end{align}
For $Q\sim \mKK \sim 1\,\meV$ this gives a suppression of roughly $(Q/\mu_{n\ell})^2\sim 10^{-24}$. 

\emphb{$\bm d$-Wave Suppression.} Near threshold the decay is pure $d$-wave. Using the parametrization of the polarizations $\varepsilon_i^{\mu\nu}=\varepsilon_i^\mu\varepsilon_i^\nu$ gives $V=\tfrac18(\varepsilon_1^\mu\varepsilon_2^\rho\varepsilon_3^\alpha Y_{\mu\rho\alpha})^2$, i.e. the square of the three-gluon amplitude. In the parent rest frame the gluon amplitude is a $p$-wave,
\begin{align}\label{eq:W}
    &\varepsilon_1^\mu\varepsilon_2^\rho\varepsilon_3^\alpha Y_{\mu\rho\alpha}=2\,\vec p\cdot\vec W,\notag\\
    &\vec W=-(\vec\varepsilon_2\!\cdot\vec\varepsilon_3)\,\vec\varepsilon_1+\frac{\vec\varepsilon_3\!\cdot\vec\varepsilon_1}{r_n}\,\vec\varepsilon_2+\frac{\vec\varepsilon_1\!\cdot\vec\varepsilon_2}{r_\ell}\,\vec\varepsilon_3,
\end{align}
with $\pm\vec p$ the daughter momenta and $\vec\varepsilon_i$ the rest-frame polarization vectors. Each momentum in $Y$ is contracted with a polarization and gives one power of $\vec p$, and the factors $1/r_{n,\ell}$ come from the time components $\varepsilon^0={\vec p\cdot\vec\varepsilon}/m$ of the moving daughters. The square of a $p$-wave contains an $s$-wave contribution proportional to $p^2\,\vec W\cdot\vec W$ and a $d$-wave contribution. Since $\vec\varepsilon_i\cdot\vec\varepsilon_i=0$,
\begin{align}\label{eq:swave}
    p^2\vec W\cdot\vec W=\frac{2p^2Qm_m}{m_nm_\ell}\Tr(\varepsilon_1\varepsilon_2\varepsilon_3).
\end{align}
Since $p^2Q\propto p^4$, this $s$-wave component is subdominant at threshold to the $d$-wave component of the amplitude which scales as $\mathcal{O}(p^2)$.

\emphb{Origin of Cancellation.} The split of the terms in Eq.~\eqref{eq:L3} can be accounted for by a change of variables. The redefinition $h=e^{H}-1$ removes $\mathcal{L}_m$ and $\mathcal{L}_\Box$, so that up to total derivatives the cubic part of Eq.~\eqref{eq:S} is
\begin{align}\label{eq:expvars}
    S_3[H]=\frac{M_5^3}{2}\int d^5x\,e^{2A}\,\mathcal{L}_V(H)
\end{align}
for all momenta and any static warped extra dimension, with $\mathcal{L}_V(H)$ the terms of $\mathcal{L}_V$ evaluated on the five-dimensional field $H(x,y)$ in place of the modes. Inserting the mode expansion, Eq.~\eqref{eq:modes}, the weight $e^{2A}$ integrates to the overlaps $g_{mn\ell}$ of Eq.~\eqref{eq:g}. The same parametrization was used to remove cubic terms with two time derivatives from the action of inflationary tensor modes~\cite{Maldacena:2002vr}; here it also removes $\mathcal{L}_m$, because that mass is from terms with derivatives with respect to $y$.

\section{Dark Dimension Implications}\label{sec:flat}
\begin{figure}[t]
\centering
\includegraphics[width=\columnwidth]{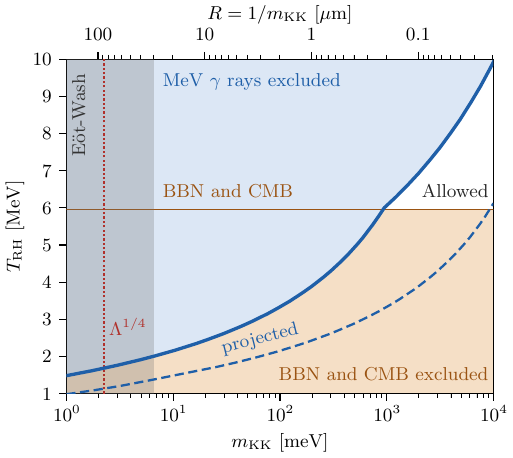}
\caption{The solid blue curve is the upper bound on $\TRH$ from MeV gamma-ray searches of the irreducible graviton population, Eq.~\eqref{eq:excess}. The horizontal line is the lower bound $\TRH\geq5.96\,\MeV$ from BBN and the CMB with BAO~\cite{Barbieri:2025moq}.  The grey strip is excluded by torsion-balance tests (for a brane at a fixed point)~\cite{Lee:2020zjt}. The dotted red line marks $\mKK=\Lambda^{1/4}$, and the dashed curve (projected) shows where the upper bound would move if proposed MeV telescopes improve bounds on gamma ray flux by a factor of 10, which is a conservative target for proposed MeV telescopes such as
AMEGO-X~\cite{Caputo:2022xpx} and e-ASTROGAM~\cite{e-ASTROGAM:2017pxr}.}
\label{fig:window}
\end{figure}

\emphb{The Cascade.} 
In the cascade scenario of Refs.~\cite{Gonzalo:2022jac,Obied:2023clp} gravitons decay down the tower to about the $\MeV$ scale by matter-radiation equality and to the keV scale today~\cite{Law-Smith:2023czn}. Equation~\eqref{eq:ratio} drastically suppresses cascades of every channel. Let us estimate the cascade rate for the example of a weakly warped extra dimension where $g_{mn\ell}$ is bounded from above because completeness of the mode profiles,
\begin{align}\label{eq:complete}
    \sum_\ell g_{mn\ell}^2=L\int dy\,e^{2A}\psi_m^2\psi_n^2=1+\mathcal{O}(\Delta A),
\end{align}
for $m\neq n$ and a warp varying by a small $\Delta A$. 

We sum Eq.~\eqref{eq:Gexact} over every mode of the flat spectrum $m_n=n\,\mKK$ with the bound of Eq.~\eqref{eq:complete} saturated. For decays losing up to $\delta n=100$ levels, the benchmark of Ref.~\cite{Gonzalo:2022jac}, and $R=0.05$-$0.2\,\mu{\rm m}$, the peak of the mass distribution moves by less than $0.1\%$ by matter-radiation equality; such an effect can be ignored.

\emphb{KK Graviton Dark Matter.} Under these assumptions, gravitons keep the mass distribution obtained at freeze-in to very high precision. They make all of the dark matter if $\TRH\sim \GeV$~\cite{Gonzalo:2022jac} for $\mKK\sim \meV$. These GeV gravitons decay into
photons with rate~\cite{Han:1998sg}
\begin{align}\label{eq:Ggg}
    \Gamma_{\gamma\gamma}=\frac{c_n^2\,m^3}{80\pi M_P^2}.
\end{align}
Such GeV gravitons would only have a lifetime of $\tau\sim \mathcal{O}(10^5\,{\rm yrs})$. Lowering $\TRH$ does not help the aim of making KK gravitons DM as this requires also decreasing $\mKK=1/R$ to values which would lead to deviations from the observed $1/r^2$ behavior of gravity~\cite{Lee:2020zjt}.

\emphb{The Irreducible Graviton Background.} Even if we relax our assumption to have gravitons be only a fraction of dark matter, an irreducible population of gravitons is produced through freeze-in via interactions with the SM plasma, i.e. Eq.~\eqref{eq:KK_graviton_coupling}. Their abundance today, assuming $\TRH\lesssim 100\,\MeV$, would be~\cite{Langhoff:2026qqg}
\begin{align}\label{eq:fh}
    \frac{\rho_h}{\rho_{\rm DM}}=1.3\times10^{-6}\Big(\frac{\TRH}{5\,\MeV}\Big)^3\left(\frac{\mKK}{5\,\meV}\right)^{-1}.
\end{align}

Searches for decaying dark matter then bound the theory itself (i.e. the existence of a dark dimension regardless of the nature of dark matter). Since the tower has many modes, we look for a spectrum of gamma rays binned in energy~\cite{Essig:2013goa}. Gravitons trace the dark matter halo and decay at the masses freeze-in gave them, producing photons of energy $m_n/2$. The $\gamma$ ray flux is
\begin{align}\label{eq:flux}
    \frac{d\Phi}{dE\,d\Omega}=\frac{r_\odot\rho_\odot\bar J}{4\pi}\sum_n\frac{2f_n e^{-\Gamma_n t_0}}{m_n}\,\Gamma_{\gamma\gamma}(m_n)\,
    \,\delta\Big(E-\frac{m_n}{2}\Big),
\end{align}
with $f_n$ the fraction of $\rho_{\rm DM}$ in level $n$ produced from freeze-in in the absence of decays which are separately handled by the factor $e^{-\Gamma_n t_0}$ with the decay rate $\Gamma_n=\tfrac94\Gamma_{\gamma\gamma}(m_n)$ and $t_0$ the age of the Universe. 

We use $\bar J=1.55$, the column of a Burkert halo over the inner-Galaxy region of the COMPTEL data~\cite{Essig:2013goa} minus that at $|b|>30^\circ$, to match the background subtraction of the data, in units of $r_\odot\rho_\odot=8.5\,{\rm kpc}\times0.3\,\GeV\,{\rm cm^{-3}}$, the distance of the Sun to the center of the galaxy times the local dark matter density. The isotropic flux collects the same decays over cosmic history which are further red-shifted. A value of $\TRH$ is excluded at a given $\mKK$ if any bin $i$ of the inner-Galaxy COMPTEL data~\cite{Essig:2013goa} or of the isotropic COMPTEL and EGRET spectra~\cite{Ajello:2008xb,Strong:2004ry} satisfies
\begin{align}\label{eq:excess}
    \frac{\Phi_i^{\rm sig}}{\Phi_i^{\rm obs}+2\sigma_i}>1,
\end{align}
with $\Phi_i$ the bin-integrated flux and $\sigma_i$ its error. CMB energy injection bounds~\cite{Slatyer:2016qyl} are weaker for $\mathcal{O}(\MeV)$ masses.

We also have a lower bound on the reheating temperature; BBN and the CMB with BAO require $\TRH\gtrsim6\,\MeV$ at $95\%$ CL. Together, the upper and lower bounds on $\TRH$ constrain the dark dimension itself: a viable $\TRH$ exists only for $R\lesssim0.2\,\mu{\rm m}$, $\mKK\gtrsim1\,\eV$ (Fig.~\ref{fig:window}).

\section{Conclusions}\label{sec:conclusions}

We have shown near threshold KK graviton decays are $d$-wave suppressed and give $\Gamma\propto p^5$ rather than the previously assumed $s$-wave scaling. This suppression prevents nearly mass-conserving cascade from transferring the high mass freeze-in population to low mass modes and excludes KK-graviton dark-matter in the approximately flat dark dimension studied here.

Even if dark matter has another origin, the irreducible freeze-in population produces MeV gamma rays that constrain $\TRH$. Combining this upper limit with $\TRH\gtrsim6\,\MeV$ from BBN and CMB+BAO constrains any viable dark dimension to have size $R\lesssim0.2\,\mu{\rm m}$, or $\mKK\gtrsim1\,\eV$. These values are far from the naive estimate $R=\mKK^{-1}\sim \Lambda^{-1/4}$. Improved MeV gamma-ray observations can further test this scenario.

\begin{acknowledgments}
\noindent 
I thank the authors of \cite{Lee:2026myr} for useful discussions which identified we had found the same decay rate suppression independently. This material is based upon work supported by the U.S. Department of Energy, Office of Science, Office of High Energy Physics of U.S. Department of Energy under grant Contract Number  DE-SC0012567 (High Energy Theory research) and Simons Foundation Investigation Awards 929255 and 929241.
\end{acknowledgments}

\bibliography{bibliography_v4}

@misc{Vafa:2024fpx,
    author = "Vafa, Cumrun",
    title = "{Swamplandish Unification of the Dark Sector}",
    eprint = "2402.00981",
    archivePrefix = "arXiv",
    primaryClass = "hep-ph",
    month = "2",
    year = "2024"
}

@article{Barbieri:2025moq,
    author = "Barbieri, Nicola and Brinckmann, Thejs and Gariazzo, Stefano and Lattanzi, Massimiliano and Pastor, Sergio and Pisanti, Ofelia",
    title = "{Current Constraints on Cosmological Scenarios with Very Low Reheating Temperatures}",
    eprint = "2501.01369",
    archivePrefix = "arXiv",
    primaryClass = "astro-ph.CO",
    doi = "10.1103/j5rj-dz1k",
    journal = "Phys. Rev. Lett.",
    volume = "135",
    number = "18",
    pages = "181003",
    year = "2025"
}

@article{Law-Smith:2023czn,
    author = "Law-Smith, Jamie A. P. and Obied, Georges and Prabhu, Anirudh and Vafa, Cumrun",
    title = "{Astrophysical constraints on decaying dark gravitons}",
    eprint = "2307.11048",
    archivePrefix = "arXiv",
    primaryClass = "hep-ph",
    doi = "10.1007/JHEP06(2024)047",
    journal = "JHEP",
    volume = "06",
    pages = "047",
    year = "2024"
}

@article{Han:1998sg,
    author = "Han, Tao and Lykken, Joseph D. and Zhang, Ren-Jie",
    title = "{On Kaluza-Klein states from large extra dimensions}",
    eprint = "hep-ph/9811350",
    archivePrefix = "arXiv",
    reportNumber = "MADPH-98-1092, FERMILAB-PUB-98-364",
    doi = "10.1103/PhysRevD.59.105006",
    journal = "Phys. Rev. D",
    volume = "59",
    pages = "105006",
    year = "1999"
}

@article{Maldacena:2002vr,
    author = "Maldacena, Juan Martin",
    title = "{Non-Gaussian features of primordial fluctuations in single field inflationary models}",
    eprint = "astro-ph/0210603",
    archivePrefix = "arXiv",
    doi = "10.1088/1126-6708/2003/05/013",
    journal = "JHEP",
    volume = "05",
    pages = "013",
    year = "2003"
}

@article{Hall:1999mk,
    author = "Hall, Lawrence J. and Tucker-Smith, David",
    title = "{Cosmological constraints on theories with large extra dimensions}",
    eprint = "hep-ph/9904267",
    archivePrefix = "arXiv",
    reportNumber = "LBNL-43091, UCB-PTH-99-14, LBL-43091",
    doi = "10.1103/PhysRevD.60.085008",
    journal = "Phys. Rev. D",
    volume = "60",
    pages = "085008",
    year = "1999"
}

@article{Lee:2020zjt,
    author = "Lee, J. G. and Adelberger, E. G. and Cook, T. S. and Fleischer, S. M. and Heckel, B. R.",
    title = "{New Test of the Gravitational $1/r^2$ Law at Separations down to 52 $\mu$m}",
    eprint = "2002.11761",
    archivePrefix = "arXiv",
    primaryClass = "hep-ex",
    doi = "10.1103/PhysRevLett.124.101101",
    journal = "Phys. Rev. Lett.",
    volume = "124",
    number = "10",
    pages = "101101",
    year = "2020"
}

@article{Dienes:2012jb,
    author = "Dienes, Keith R. and Thomas, Brooks",
    title = "{Phenomenological Constraints on Axion Models of Dynamical Dark Matter}",
    eprint = "1203.1923",
    archivePrefix = "arXiv",
    primaryClass = "hep-ph",
    reportNumber = "UH511-1189-12",
    doi = "10.1103/PhysRevD.86.055013",
    journal = "Phys. Rev. D",
    volume = "86",
    pages = "055013",
    year = "2012"
}

@article{Obied:2023clp,
    author = "Obied, Georges and Dvorkin, Cora and Gonzalo, Eduardo and Vafa, Cumrun",
    title = "{Dark dimension and decaying dark matter gravitons}",
    eprint = "2311.05318",
    archivePrefix = "arXiv",
    primaryClass = "astro-ph.CO",
    doi = "10.1103/PhysRevD.109.063540",
    journal = "Phys. Rev. D",
    volume = "109",
    number = "6",
    pages = "063540",
    year = "2024"
}

@article{Gonzalo:2022jac,
    author = "Gonzalo, Eduardo and Montero, Miguel and Obied, Georges and Vafa, Cumrun",
    title = "{Dark dimension gravitons as dark matter}",
    eprint = "2209.09249",
    archivePrefix = "arXiv",
    primaryClass = "hep-ph",
    doi = "10.1007/JHEP11(2023)109",
    journal = "JHEP",
    volume = "11",
    pages = "109",
    year = "2023"
}

@article{Giudice:1998ck,
    author = "Giudice, Gian F. and Rattazzi, Riccardo and Wells, James D.",
    title = "{Quantum gravity and extra dimensions at high-energy colliders}",
    eprint = "hep-ph/9811291",
    archivePrefix = "arXiv",
    reportNumber = "CERN-TH-98-354",
    doi = "10.1016/S0550-3213(99)00044-9",
    journal = "Nucl. Phys. B",
    volume = "544",
    pages = "3--38",
    year = "1999"
}

@article{Montero:2022prj,
    author = "Montero, Miguel and Vafa, Cumrun and Valenzuela, Irene",
    title = "{The dark dimension and the Swampland}",
    eprint = "2205.12293",
    archivePrefix = "arXiv",
    primaryClass = "hep-th",
    doi = "10.1007/JHEP02(2023)022",
    journal = "JHEP",
    volume = "02",
    pages = "022",
    year = "2023"
}

@article{Langhoff:2022bij,
    author = "Langhoff, Kevin and Outmezguine, Nadav Joseph and Rodd, Nicholas L.",
    title = "{Irreducible Axion Background}",
    eprint = "2209.06216",
    archivePrefix = "arXiv",
    primaryClass = "hep-ph",
    reportNumber = "CERN-TH-2022-148",
    doi = "10.1103/PhysRevLett.129.241101",
    journal = "Phys. Rev. Lett.",
    volume = "129",
    number = "24",
    pages = "241101",
    year = "2022"
}

@article{Dienes:2011ja,
    author = "Dienes, Keith R. and Thomas, Brooks",
    title = "{Dynamical Dark Matter: I. Theoretical Overview}",
    eprint = "1106.4546",
    archivePrefix = "arXiv",
    primaryClass = "hep-ph",
    doi = "10.1103/PhysRevD.85.083523",
    journal = "Phys. Rev. D",
    volume = "85",
    pages = "083523",
    year = "2012"
}

@article{Dienes:2011sa,
    author = "Dienes, Keith R. and Thomas, Brooks",
    title = "{Dynamical Dark Matter: II. An Explicit Model}",
    eprint = "1107.0721",
    archivePrefix = "arXiv",
    primaryClass = "hep-ph",
    doi = "10.1103/PhysRevD.85.083524",
    journal = "Phys. Rev. D",
    volume = "85",
    pages = "083524",
    year = "2012"
}

@article{Ooguri:2006in,
    author = "Ooguri, Hirosi and Vafa, Cumrun",
    title = "{On the Geometry of the String Landscape and the Swampland}",
    eprint = "hep-th/0605264",
    archivePrefix = "arXiv",
    reportNumber = "CALT-68-2600, HUTP-06-A017",
    doi = "10.1016/j.nuclphysb.2006.10.033",
    journal = "Nucl. Phys. B",
    volume = "766",
    pages = "21--33",
    year = "2007"
}

@article{Lee:2019xtm,
    author = "Lee, Seung-Joo and Lerche, Wolfgang and Weigand, Timo",
    title = "{Emergent strings, duality and weak coupling limits for two-form fields}",
    eprint = "1904.06344",
    archivePrefix = "arXiv",
    primaryClass = "hep-th",
    reportNumber = "CERN-TH-2019-044",
    doi = "10.1007/JHEP02(2022)096",
    journal = "JHEP",
    volume = "02",
    pages = "096",
    year = "2022"
}

@article{Lee:2019wij,
    author = "Lee, Seung-Joo and Lerche, Wolfgang and Weigand, Timo",
    title = "{Emergent strings from infinite distance limits}",
    eprint = "1910.01135",
    archivePrefix = "arXiv",
    primaryClass = "hep-th",
    reportNumber = "CERN-TH-2019-159",
    doi = "10.1007/JHEP02(2022)190",
    journal = "JHEP",
    volume = "02",
    pages = "190",
    year = "2022"
}

@article{Lust:2019zwm,
    author = {L{\"u}st, Dieter and Palti, Eran and Vafa, Cumrun},
    title = "{AdS and the Swampland}",
    eprint = "1906.05225",
    archivePrefix = "arXiv",
    primaryClass = "hep-th",
    doi = "10.1016/j.physletb.2019.134867",
    journal = "Phys. Lett. B",
    volume = "797",
    pages = "134867",
    year = "2019"
}

@misc{Obied:2018sgi,
    author = "Obied, Georges and Ooguri, Hirosi and Spodyneiko, Lev and Vafa, Cumrun",
    title = "{De Sitter Space and the Swampland}",
    eprint = "1806.08362",
    archivePrefix = "arXiv",
    primaryClass = "hep-th",
    reportNumber = "CALT-TH-2018-020, IPMU18-0100",
    month = "6",
    year = "2018"
}

@article{Bedroya:2019snp,
    author = "Bedroya, Alek and Vafa, Cumrun",
    title = "{Trans-Planckian Censorship and the Swampland}",
    eprint = "1909.11063",
    archivePrefix = "arXiv",
    primaryClass = "hep-th",
    doi = "10.1007/JHEP09(2020)123",
    journal = "JHEP",
    volume = "09",
    pages = "123",
    year = "2020"
}

@article{Mohapatra:2003ah,
    author = "Mohapatra, R. N. and Nussinov, S. and Perez-Lorenzana, Abdel",
    title = "{Large extra dimensions and decaying $K K$ recurrences}",
    eprint = "hep-ph/0308051",
    archivePrefix = "arXiv",
    reportNumber = "UMD-PP-03-065",
    doi = "10.1103/PhysRevD.68.116001",
    journal = "Phys. Rev. D",
    volume = "68",
    pages = "116001",
    year = "2003"
}

@article{Bern:2008qj,
    author = "Bern, Z. and Carrasco, J. J. M. and Johansson, Henrik",
    title = "{New Relations for Gauge-Theory Amplitudes}",
    eprint = "0805.3993",
    archivePrefix = "arXiv",
    primaryClass = "hep-ph",
    reportNumber = "UCLA-07-TEP-15",
    doi = "10.1103/PhysRevD.78.085011",
    journal = "Phys. Rev. D",
    volume = "78",
    pages = "085011",
    year = "2008"
}

@article{Bern:2019prr,
    author = "Bern, Zvi and Carrasco, John Joseph and Chiodaroli, Marco and Johansson, Henrik and Roiban, Radu",
    title = "{The duality between color and kinematics and its applications}",
    eprint = "1909.01358",
    archivePrefix = "arXiv",
    primaryClass = "hep-th",
    reportNumber = "CERN-TH-2019-135, UCLA/TEP/2019/104, NUHEP-TH/19-11, UUITP-35/19,
  NORDITA 2019-079",
    doi = "10.1088/1751-8121/ad5fd0",
    journal = "J. Phys. A",
    volume = "57",
    number = "33",
    pages = "333002",
    year = "2024"
}

@article{Chivukula:2024nzt, author = {Chivukula, R. Sekhar and Gill, Joshua A. and Mohan, Kirtimaan A. and Sanamyan, George and Sengupta, Dipan and Simmons, Elizabeth H. and Wang, Xing}, title = {{Limits on Kaluza-Klein Portal Dark Matter Models}}, eprint = {2411.02509}, archivePrefix = {arXiv}, journal = {Phys. Rev. D}, volume = {111}, pages = {075030}, year = {2025}, doi = {10.1103/PhysRevD.111.075030}}

@article{Kawai:1985xq,
    author = "Kawai, H. and Lewellen, D. C. and Tye, S. H. H.",
    title = "{A Relation Between Tree Amplitudes of Closed and Open Strings}",
    reportNumber = "CLNS-85/667",
    doi = "10.1016/0550-3213(86)90362-7",
    journal = "Nucl. Phys. B",
    volume = "269",
    pages = "1--23",
    year = "1986"
}

@article{deGiorgi:2021xvm,
    author = "de Giorgi, Arturo and Vogl, Stefan",
    title = "{Dark matter interacting via a massive spin-2 mediator in warped extra-dimensions}",
    eprint = "2105.06794",
    archivePrefix = "arXiv",
    primaryClass = "hep-ph",
    doi = "10.1007/JHEP11(2021)036",
    journal = "JHEP",
    volume = "11",
    pages = "036",
    year = "2021"
}

@article{Slatyer:2016qyl,
    author = "Slatyer, Tracy R. and Wu, Chih-Liang",
    title = "{General Constraints on Dark Matter Decay from the Cosmic Microwave Background}",
    eprint = "1610.06933",
    archivePrefix = "arXiv",
    primaryClass = "astro-ph.CO",
    reportNumber = "MIT-CTP-4842",
    doi = "10.1103/PhysRevD.95.023010",
    journal = "Phys. Rev. D",
    volume = "95",
    number = "2",
    pages = "023010",
    year = "2017"
}

@article{Essig:2013goa, author = {Essig, Rouven and Kuflik, Eric and McDermott, Samuel D. and Volansky, Tomer and Zurek, Kathryn M.}, title = {{Constraining Light Dark Matter with Diffuse X-Ray and Gamma-Ray Observations}}, eprint = {1309.4091}, archivePrefix = {arXiv}, journal = {JHEP}, volume = {11}, pages = {193}, year = {2013}, doi = {10.1007/JHEP11(2013)193}}

@article{Ajello:2008xb,
    author = "Ajello, M. and others",
    title = "{Cosmic X-ray background and Earth albedo Spectra with Swift/BAT}",
    eprint = "0808.3377",
    archivePrefix = "arXiv",
    primaryClass = "astro-ph",
    doi = "10.1086/592595",
    journal = "Astrophys. J.",
    volume = "689",
    pages = "666",
    year = "2008"
}

@article{Strong:2004ry,
    author = "Strong, A. W. and Moskalenko, I. V. and Reimer, O.",
    title = "{A new determination of the extragalactic diffuse gamma-ray background from egret data}",
    eprint = "astro-ph/0405441",
    archivePrefix = "arXiv",
    doi = "10.1086/423196",
    journal = "Astrophys. J.",
    volume = "613",
    pages = "956--961",
    year = "2004"
}

@misc{Langhoff:2026qqg,
    author = "Langhoff, Kevin and Ramos, Maria and Reig, Mario",
    title = "{The Dark Dimension meets the Axiverse}",
    eprint = "2607.27314",
    archivePrefix = "arXiv",
    primaryClass = "hep-ph",
    reportNumber = "CERN-TH-2026-169, MIT-CTP/6073",
    month = "7",
    year = "2026"
}

@misc{Lee:2026myr,
    author = "Lee, Vincent S. H. and Randall, Lisa and Riojas, Marcos",
    title = "{A Universal Kaluza-Klein Graviton Cascade Rate}",
    eprint = "2609.36234",
    archivePrefix = "arXiv",
    primaryClass = "hep-th",
    month = "9",
    year = "2026"
}

@article{Hannestad:2001nq,
    author = "Hannestad, Steen",
    title = "{Strong constraint on large extra dimensions from cosmology}",
    eprint = "hep-ph/0102290",
    archivePrefix = "arXiv",
    doi = "10.1103/PhysRevD.64.023515",
    journal = "Phys. Rev. D",
    volume = "64",
    pages = "023515",
    year = "2001"
}

@article{Caputo:2022xpx,
    author = "Caputo, Regina and others",
    title = "{All-sky Medium Energy Gamma-ray Observatory eXplorer mission concept}",
    eprint = "2208.04990",
    archivePrefix = "arXiv",
    primaryClass = "astro-ph.IM",
    doi = "10.1117/1.JATIS.8.4.044003",
    journal = "J. Astron. Telesc. Instrum. Syst.",
    volume = "8",
    number = "4",
    pages = "044003",
    year = "2022"
}

@article{e-ASTROGAM:2017pxr,
    author = "Tavani, M. and others",
    editor = "De Angelis, A. and Tatischeff, V. and Grenier, I. A. and McEnery, J. and Mallamaci, M.",
    collaboration = "e-ASTROGAM",
    title = "{Science with e-ASTROGAM: A space mission for MeV{\textendash}GeV gamma-ray astrophysics}",
    eprint = "1711.01265",
    archivePrefix = "arXiv",
    primaryClass = "astro-ph.HE",
    doi = "10.1016/j.jheap.2018.07.001",
    journal = "JHEAp",
    volume = "19",
    pages = "1--106",
    year = "2018"
}
\end{document}